\documentclass[journal=jctcce, manuscript=article, 
                        layout=twocolumn 
                        ]{achemso}

\usepackage{chemformula} 
\usepackage[T1]{fontenc} 
\usepackage{amsmath, amsfonts, amssymb, amsthm} 
\usepackage{bm} 
\usepackage{hyperref} 
\usepackage{nameref}  

\usepackage{soul}  
\soulregister\ref7
\soulregister\cite7
\soulregister\ce7
\soulregister\textrm7

\author{Leonard P. Heinz}
\affiliation[MPIBPC]{Department of Theoretical and Computational Biophysics, Max-Planck Institute for Biophysical Chemistry, G{\"o}ttingen, Germany}
\email{lheinz@gwdg.de}

\author{Helmut Grubm{\"u}ller}
\affiliation[MPIBPC]{Department of Theoretical and Computational Biophysics, Max-Planck Institute for Biophysical Chemistry, G{\"o}ttingen, Germany}
\email{hgrubmu@gwdg.de}

\newcommand{\JmolK}{J$\cdot$mol$^{-1}\cdot$K$^{-1}$}
\newcommand{\kJmol}{kJ$\cdot$mol$^{-1}$}

\title{Computing spatially resolved rotational hydration entropies from atomistic simulations}

\abbreviations{}

\begin{document}
\setcounter{secnumdepth}{3}

\begin{abstract}
    For a first principles understanding of macromolecular processes, a quantitative understanding of the underlying free energy landscape and in particular its entropy contribution is crucial. The stability of biomolecules, such as proteins, is governed by the hydrophobic effect, which arises from competing enthalpic and entropic contributions to the free energy of the solvent shell. While the statistical mechanics of liquids, as well as molecular dynamics simulations have provided much insight, solvation shell entropies remain notoriously difficult to calculate, especially when spatial resolution is required. Here, we present a method that allows for the computation of spatially resolved rotational solvent entropies via a non-parametric $k$-nearest-neighbor density estimator. We validated our method using analytic test distributions and applied it to atomistic simulations of a water box. With an accuracy of better than $9.6\,$\%, the obtained spatial resolution should shed new light on the hydrophobic effect and the thermodynamics of solvation in general. 
\end{abstract}

\section{Introduction}

Competing enthalpic and entropic contributions to the solvation free energies give rise to the hydrophobic effect\cite{Ben_Naim_1975_hydrophobic_effect}, which is vital for protein function and folding\cite{Chandler_2005, Berne_2009_hydrophobicity_review, Dias_2010_cold_denaturation}. Despite extensive theoretical work\cite{Ben_Naim_1975_hydrophobic_effect, Aveblj_2017_hydrophobic_effect}, a quantitative understanding of the hydrophobic effect particularly at heterogeneous surfaces, such as of proteins and mixed bilayers, remains elusive.

Because surface-water shows a significantly altered behaviour compared to bulk\cite{Cheng_1998,Tarek_2000}, it is essential for our understanding of the thermodynamics and energetics of protein solvation to better characterize, e.g., the relative contributions by different solvation shells or the effect of individual protein side chains on the solvent. Molecular dynamics (MD) simulations describe the hydrophobic effect at an atomic level\cite{Sugita_1999_replica_exchange, Devrjes_2004_bilayer_assembly}, but a deeper understanding of the molecular driving forces requires a quantitative and spatially resolved picture of solvation shell thermodynamics, which poses considerable challenges.

Methods like thermodynamic integration (TI)\cite{Kirkwood_1935_TI,Peter_2004_TI} allow for the calculation of solvation entropies based on MD simulations, but the lack of a spatial resolution precludes detailed analysis of how local features of the solvent-surface interface contribute and interact. Various order parameters\cite{Errington_2001_orderparam, Errington_2002_orderparam, Yan_2007_orderparam, Godec_2011, Heyden_2019} assess both the local translational and the local rotational order of water molecules but yield only a qualitative picture of the thermodynamic entropy.

Here, we limit our analysis to absolute rotational water entropies and present a method to reach a spatial resolution from atomistic simulations or Monte Carlo ensembles. Our method employs a mutual information expansion (MIE) to calculate the total entropy of $N$ water molecules based on the contributions of each molecule individually and the entropy loss due to correlations between molecule pairs and triples. A similar approach was taken by, e.g., the grid inhomogeneous solvation theory (GIST)\cite{Wallace_1987_IST,Baranyai_1989_IST, Lazaridis_1998A_IST, Lazaridis_1998B_IST,Nguyen_2012_GIST,Nguyen_2012_GIST_erratum, Nguyen_2015_GIST_KNN}. Rather than considering entropic contributions by correlations between individual molecules directly, GIST calculates discretised correlation-integrals within voxels, which causes severe sampling problems for higher order correlations. 3D-2-Phase-Thermodynamics (3D-2PT)\cite{Lin_2003_2pt,Lin_2010_2pt,Persson_2017_3d2pt} also uses voxels and approximates the system as a superposition of gas-like and solid-like components. Likewise, the Grid Cell Theory (GCT)\cite{GCT} includes free energies and enthalpies, but it approximates rotational water correlation terms using a generalized Pauling's residual ice entropy model\cite{Pauling_ice_entropy_1935, Henchman_2010}. Here, we address these correlations directly, convergence of which is challenging, as they require sampling and density estimates in high-dimensional configuration spaces. 

In our approach, all MIE terms were calculated using a $k$-nearest-neighbor (kNN) density estimator, typically used in Euclidean spaces\cite{Kozachenko_1987_kNN, Singh_2003_kNN, Kraskov_2004_kNN}, which we modified and optimized for $SO(3)^n$, the cartesian products of the group of rotations. We considered different metrices for the $k$-nearest neighbors in $SO(3)^n$, determined an optimal $k$-value, and provide a computationally efficient framework for rotational entropy calculation. 

For easier notation, will develop our method for water molecules, although it is general and applicable to any system with rotational degrees of freedom.

In the following sections, we will first provide the conceptual foundation and then describe our rotation entropy approach. Subsequently, we will apply it to analytical test distributions, as well as to MD water boxes.  

\section{Theory}
\label{sec:theory}

\subsection{Absolute entropy}

Separating the entropy of water into rotational and translational contributions yields
\begin{equation*}
    S_\text{total} = S_\text{rotation} + S_\text{translation} - I_\text{corr},
    \label{eq:entropy_separation}
\end{equation*}
where $S_\text{rotation}$ is the entropy of the phase space distribution after projection onto the rotational degrees of freedom; $S_\text{translation}$, respectively, is the entropy arising from translational degrees of freedom; and the mutual information (MI) term $I_\text{corr}$ quantifies the correlations between translation and rotation. In this paper, we focus on the rotational contribution $S_\text{rotation}$.

Note that some authors\cite{Lazaridis_1996, Nguyen_2012_GIST, Nguyen_2015_GIST_KNN} define the rotational entropy as a conditional entropy, in which case it includes the MI term $-I_\text{corr}$.

Let the rotation of $N$ water molecules of the simulation system be described by the Hamiltonian $\mathcal{H}(\{ \bm{L}_i, \bm{\omega}_i \}) = \mathcal{T}(\{ \bm{L}_i \}) + \mathcal{V}(\{ \bm{\omega}_i \})$, with angular momenta $\bm{L}_i$, orientations $\bm{\omega}_i \in SO(3)$, the kinetic energy $\mathcal{T}$, and the potential energy $\mathcal{V}$, typically described by a molecular mechanics force field. The total entropy is
\begin{equation*}
    S_\text{rotation} = - k_B \int \frac{d\bm{L}^N d\bm{\omega}^N}{h^{3N}} \varrho \log \varrho,
\end{equation*}
with the Boltzmann constant $k_B$, Planck's constant $h$, and the normalized and dimensionless phase space density $\varrho = Z^{-1} \exp{\left[ -\frac{\mathcal{H}}{k_BT} \right]} = \varrho_\mathcal{T} \varrho_\mathcal{V}$, with $\varrho_\mathcal{T} = Z_\mathcal{T}^{-1} \exp{\left[ -\frac{\mathcal{T}}{k_BT} \right]}$, $\varrho_\mathcal{V} = Z_\mathcal{V}^{-1} \exp{\left[ -\frac{\mathcal{V}}{k_BT} \right]}$, and the partiton function $Z=Z_\mathcal{T} Z_\mathcal{V}$. Because $\varrho$ factorizes, the entropy can be split into a kinetic and a configurational term
\begin{align*}
    S_\text{rotation}    &= 
    \!\begin{aligned}[t]
        &- k_B \int \frac{d\bm{L}^N}{h_\mathcal{T}^{3N}} \varrho_\mathcal{T} \log \varrho_\mathcal{T} \\
        &- k_B \int \frac{d\bm{\omega}^N}{h_\mathcal{V}^{3N}} \varrho_\mathcal{V} \log \varrho_\mathcal{V}
    \end{aligned} \\
                    &=
    \!\begin{aligned}[t]
        &\underbrace{ \frac{3Nk_B}{2} \log \left[ \frac{2\pi e k_B T}{h_\mathcal{T}^2} \prod_{i=1}^3 I_i^\frac{1}{3} \right] }_{S_\text{kin}} \\
        & \underbrace{- k_B \int \frac{d\bm{\omega}^N}{h_\mathcal{V}^{3N}} \varrho_\mathcal{V} \log \varrho_\mathcal{V} }_{S_\text{conf}},
    \end{aligned}
\end{align*}
where $h_\mathcal{T}>0$ is arbitrary, $h_\mathcal{V} = h/h_\mathcal{T}$, and $I_i$ are the eigenvalues of the moment-of-inertia tensor of a water molecule.

Because $S_\text{kin}$ can be solved analytically, the challenge is to estimate $S_\text{conf}$. 

\subsection{Entropy estimation}

Because the rotational entropy integral in $3N$ dimensions usually cannot be computed directly, we used a truncated mutual information expansion\cite{Matsuda_2000,Hnizdo_2007,Hnizdo_2008,Fengler_thesis} (see section~\ref{subsubsec:mutual_information_expansion}) to expand the full high-dimensional integral into multiple low-dimensional integrals over marginal distributions, which can be calculated numerically, similarly to the Inhomogeneous Solvation Theory (IST)\cite{Lazaridis_1998A_IST,Lazaridis_1998B_IST,Killian_2007}, underlying GIST. To obtain these marginal entropies, a $k$-nearest-neighbor estimator (see section~\ref{subsubsec:kNN_entropy_estimation}), which estimates the density at each sample point by finding the $k$ closest neighboring sample points and dividing by the volume of a ball that encloses the points, was used. Here, the orientations of $N$ water molecules in $n_f$ different samples, e.g., frames of a computer simulation trajectory, were represented by a series of quaternions (see section~\ref{subsubsec:parametrization_of_orientations}) $\{ \bm{q}_{i,1}, \hdots, \bm{q}_{i,n_f} \}$ with $i=1,\hdots,N$. We then defined suitable distance metrics, as required by the kNN algorithm, which are not trivial in curved spaces of rotations $SO(3)^n$ (see section~\ref{subsubsec:distances_in_SO3}), and then calculated the volumes of balls, as induced by the metrics (see section~\ref{subsubsec:volumes_of_balls_in_SO3}). We finally present a computationally efficient framework that allows finding $k$ neighbors to each sample point (see section~\ref{subsec:nearest_neighbor_search}).

\subsubsection{Mutual information expansion}
\label{subsubsec:mutual_information_expansion}

\begin{figure*}
    \centering
    \includegraphics[scale=1]{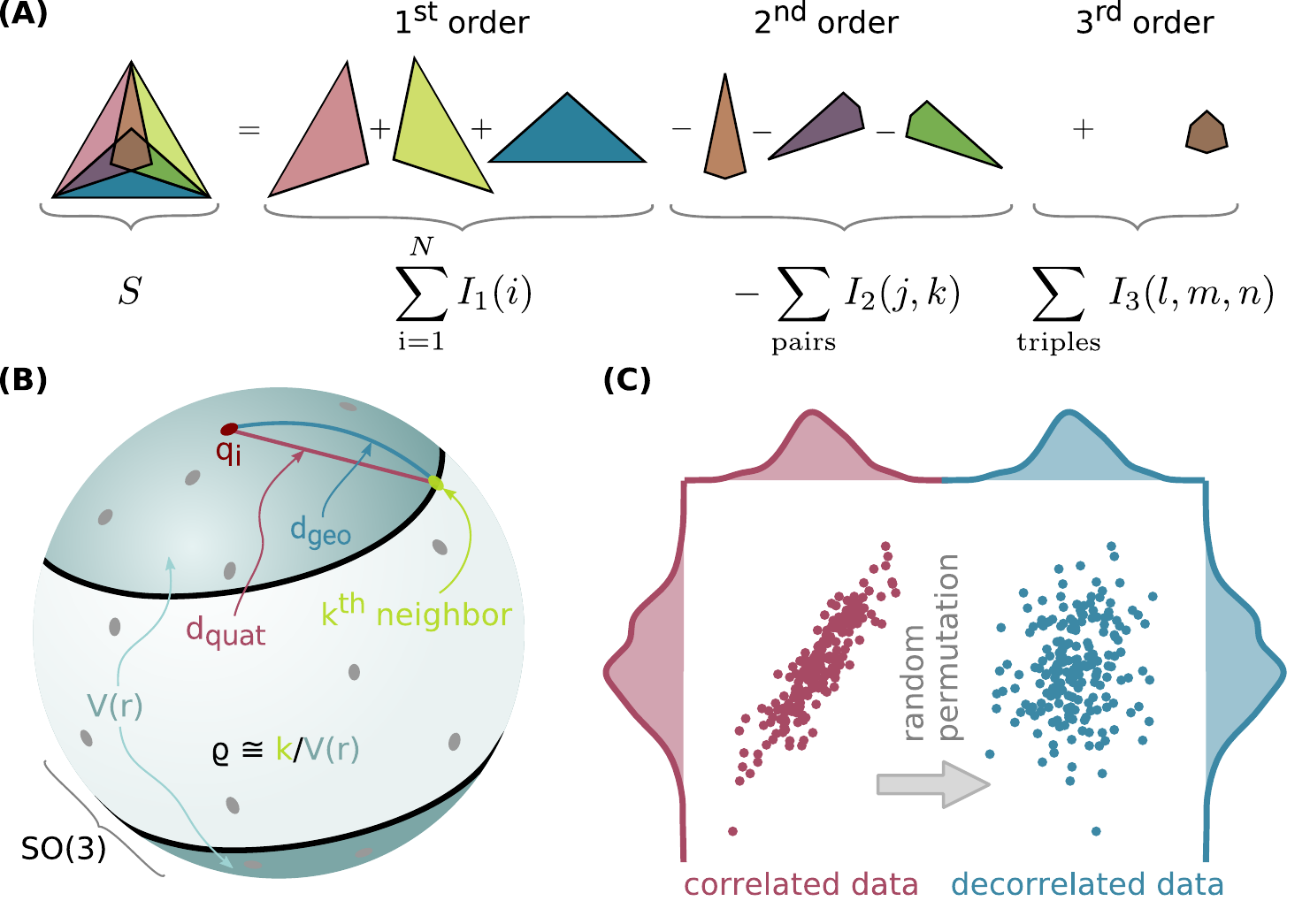}
    \caption{\textbf{(A)} Mutual information expansion illustrated for the entropy-breakdown of 3 particles. \textbf{(B)} Sketch of density estimation on $SO(3)$ (here represented as a 2-sphere). Each dot on the sphere represents an orientation. For each point $\bm{x}_i$, the $k$th neighbor according to a distance metric (e.g., $d_\text{quat}$ or $d_\text{geo}$) is found. The density is estimated via the volume $V(r)$ of a ball with radius $r=d(\cdot,\cdot)$. \textbf{(C)} Visualization of the fill mode approach: A correlated dataset is shown on the left hand side. The identical data is decorrelated by applying a random permutation along one axis, as shown on the right. The entropy of the decorrelated data is the sum of both "marginal entropies".}
    \label{fig:explanation_fig}
\end{figure*}

Figure~\ref{fig:explanation_fig}A shows an example of an entropy expansion into mutual information (MI) terms of a system containing three subsystems, such as three water molecules, in a Venn diagram: The full entropy ($S$) is expanded into MI terms ($I_m$), of which the first term represents the entropies of each molecule individually and the further terms are correlation terms of 2nd and 3rd order, respectively,
\begin{subequations}
\label{eq:def_MI1_to_3}
\begin{align}
    I_1(i)      &=S(i) \\
    I_2(j,k)    &=S(j)+S(k)-S(j,k) \label{eq:def_MI2} \\
    I_3(l,m,n)  &=
    \!\begin{aligned}[t]
        &S(l)+S(m)+S(n) \\
        &-S(l,m)-S(l,n)-S(m,n)\\
        &+S(l,m,n).
    \end{aligned}
    \label{eq:def_MI3}
\end{align}
\end{subequations}
In this notation, $S(\gamma_1,\hdots, \gamma_m)$ is the entropy of the marginal distribution with respect to molecules with indices $\gamma_1,\hdots, \gamma_m$.

For $N$ water molecules, the expansion consists of $N$ MI orders, of which the $m$th term involves $(3 m)$-dimensional integrals and takes all possible $m$-molecule correlations into account. Approximating the full entropy by a truncated expansion thus leads to lower dimensional integrals, which can be better sampled. Although there is no guarantee that truncated orders are small and can be neglected, it has been shown that a truncated expansion provides accurate entropy estimates if the correlations are short ranged\cite{Rubi_2017}, as for water in physiological conditions.

Here, we took up to $3$-molecule-correlations into account by truncating after the 3rd order, hence
\begin{equation}
    \begin{aligned}
        S   &\approx \sum_{i=1}^N I_1(i) \\
            &- \sum_{(j,k)\in \text{pairs}} I_2(j,k) \\
            &+ \sum_{(l,m,n)\in \text{triples}} I_3(l,m,n),
    \end{aligned}
    \label{eq:MIE_trunc}
\end{equation}
where the 1st order includes the kinetic entropy contribution and a correction of $-N \log 2$, due to the two-fold symmetry of the water molecule. The three terms are akin to the terms in IST\cite{Nguyen_2015_GIST_KNN}. In fact, closer analysis shows that in the thermodynamic limit, the 2nd and 3rd order terms in the IST-expansion\cite{Wallace_1987_IST,Baranyai_1989_IST,Lazaridis_1998A_IST} of the molar entropy converge towards the respective terms in eq \ref{eq:MIE_trunc}.  

\subsubsection{kNN entropy estimation}
\label{subsubsec:kNN_entropy_estimation}

To evaluate eq \ref{eq:MIE_trunc} from a given sample of orientations $\{ \bm{q}_1, \hdots, \bm{q}_{n_f} \}_i$ with $i=1,\hdots,N$, the marginal entropies from eq \ref{eq:def_MI1_to_3} are calculated using a kNN entropy estimator\cite{Kozachenko_1987_kNN, Tsybakov_1996_kNN,Singh_2003_kNN,Kraskov_2004_kNN,Evans_2008_kNN}. For $SO(3)^1$, the $k$th nearest neighbor with respect to the sample point $\bm{q}_i$ is defined by a metric $d(\bm{q}_i, \bm{q}_j)$ (see Figure~\ref{fig:explanation_fig}B), and $\varrho(\bm{q_i})$ is estimated as $(n_f-1)^{-1} k/V(r_{i,k})$, where $k$ is a fixed integer, $V(r_{i,k})$ is the volume of a ball with radius $r_{i,k}$, the distance between $\bm{q}_i$ and its $k$th neighbor, and $(n_f-1)^{-1}$ is a normalization constant. Results for $SO(3)^2$ and $SO(3)^3$ are obtained by generalizing the metric $d$ and the volume $V(r_{i,k})$ to higher dimensions. The choice of metrices, on which the results may depend for finite sampling, and their corresponding volumes in $SO(3)^n$ will be discussed in section~\ref{subsubsec:distances_in_SO3} and section~\ref{subsubsec:volumes_of_balls_in_SO3}. The entropy is
\begin{align*}
    S &= -\langle \log \varrho \rangle \\
      &\approx - \frac{1}{n_f} \sum_{i=1}^{n_f} \log \left( \frac{k}{(n_f-1)V(r_{i,k})} \right) - \gamma_k,
\end{align*}
where $\gamma_k = \psi(k) - \log k$ is a correction which accounts for the bias introduced by the $k$th neighbors being, by definition, on the edges of the balls\cite{Kraskov_2004_kNN}. $\psi$ is the digamma function.

Because eqs \ref{eq:def_MI2} and \ref{eq:def_MI3} are sums and differences of integrals of different dimensionalities, biases are introduced: With increasing dimensionality and thus reduced sampling, the kNN estimator yields increasingly smoothed versions of the underlying true distributions. The estimator therefore overestimates entropies of distributions with higher-dimensional supports more than of those defined in lower-dimensional spaces, resulting in biases if entropies of different dimensionality are added or subtracted. To overcome this problem, the sampling space is expanded to equal dimensionality by using fill modes\cite{Hensen_2010_min_coupl_subsp,Fengler_thesis}. $I_2$, defined in eq \ref{eq:def_MI2} as the sum of integrals in $SO(3)^1$ and $SO(3)^2$, can be rewritten as a sum of two $SO(3)^2$ integrals
\begin{equation*}
    I_2(j,k) = S(j,\hat{k}) - S(j,k)
    \label{eq:def_MI2_fill}
\end{equation*}
if the corresponding joint distribution $\varrho(j,\hat{k})$ factorizes to $\varrho(j) \varrho(\hat{k}) = \varrho(j) \varrho(k)$. To achieve statistical independence, the sample points corresponding to index $k$ were subjected to a random permutation $\{ \bm{q}_{\hat{k},1}, \hdots, \bm{q}_{\hat{k},n_f} \} = \text{perm} \{ \bm{q}_{k,1}, \hdots, \bm{q}_{k,n_f} \}$, which decorrelates $\{ \bm{q}_{j,1}, \hdots, \bm{q}_{j,n_f} \}$ and $\{ \bm{q}_{k,1}, \hdots, \bm{q}_{k,n_f} \}$, but leaves the marginal distributions unchanged, as sketched in Figure~\ref{fig:explanation_fig}C. The joint entropy $S(j, \hat{k})$ is thus the sum of the initial marginal entropies $S(j)+S(k)$. 

Similarly, the 3rd order MI term reads
\begin{align*}
    I_3(l,m,n)  &= 2S(\hat{l},\hat{m},\hat{n}) \\
                &- S(l,m,\hat{n}) - S(l,\hat{m},n) - S(\hat{l},m,n) \\
                &+ S(l,m,n).
\end{align*}

\subsubsection{Parametrization of orientations}
\label{subsubsec:parametrization_of_orientations}

From different parametrizations of orientations in 3D-space, such as Euler angles, Tait-Bryan angles, Hopf coordinates\cite{Hopf_1964, Jain_2010_Hopf}, and spherical coordinates, we used quaternions\cite{Karney_2007_quaternions}, which, contrary to most other charts of $SO(3)$, do not suffer from Gimbal lock. They are defined as $\bm{q} = (q_1, q_2, q_3, q_4)^\intercal = \pm (\cos \frac{\theta}{2}, \bm{u}^\intercal \sin \frac{\theta}{2})^\intercal$, where $\bm{u}$ and $\theta$ are a normalized rotation axis and a rotation angle, respectively. $\bm{q}$ can thus be interpreted as an element of the 3-sphere, i.e., $\| \bm{q} \|_2 = 1$. Because there is a one-to-one mapping of the 3-sphere to the Special Unitary group $SU(2)$, which in turn provides a double-covering of $SO(3)$, each orientation is described by two equivalent quaternions, which differ only by a sign\cite{Huynh_2009}.

\subsubsection{Choice of metrics in $SO(3)^n$}
\label{subsubsec:distances_in_SO3}

We next considered the proper choice of metrics in $SO(3)^n$. At first sight, one might think that, of many possible metrics in $SO(3)$\cite{Huynh_2009, Huggins_2014}, only one, e.g., the geodesic metric $d_\text{geo}(\bm{q}_1, \bm{q}_2) = \arccos{(| \bm{q}_1 \cdot \bm{q}_2 |)}$ shown in Figure~\ref{fig:explanation_fig}B, yields the correct entropy. However, in the limit of infinite sampling, kNN entropy estimation with any metric is possible if used with its induced ball-volumes (see section~\ref{subsubsec:volumes_of_balls_in_SO3})\cite{Singh_2016_kNN}. Our choice was therefore guided by the speed of convergence and computational efficiency. 

We chose the quaternion metric\cite{Ravani_1983_quaternionmetric, Huynh_2009}
\begin{equation*}
    d_\text{quat}(\bm{q}_1, \bm{q}_2) = \text{min} \{ \| \bm{q}_1 - \bm{q}_2 \|_2,  \| \bm{q}_1 + \bm{q}_2 \|_2 \},
    \label{eq:def_quat}
\end{equation*}
sketched in Figure~\ref{fig:explanation_fig}B, which defines a metric between two rotations as the minimum Euclidean distance between unit quaternions, taking the sign ambiguity into account. In $SO(3)$, the quaternion metric and the more natural geodesic metric $d_\text{quat}$ yield identical nearest neighbors. They are functionally equivalent because a positive continuous strictly increasing function $h$, such that $h \circ d_\text{geo} = d_\text{quat}$ (and vice versa), exists\cite{Huynh_2009}. $d_\text{quat}$ does not require evaluation of the inverse cosine function and thus is computationally more efficient; it was therefore preferred over $d_\text{geo}$.

Metrices in $SO(3)^2$ and $SO(3)^3$ were obtained by combining $d_\text{quat}$ with the Euclidean norms in $\mathbb{R}^2$ and $\mathbb{R}^3$, respectively,
\begin{align*}
    d_{\text{quat}^n}((\bm{q}_{1,1}, \hdots, \bm{q}_{1,n}), (\bm{q}_{2,1}, \hdots, \bm{q}_{2,n}) ) \\
    = \sqrt{ \sum_{i=1}^n  d_\text{quat}(\bm{q}_{1,i}, \bm{q}_{2,i})^2},
\end{align*}
with $(\bm{q}_{i,1}, \hdots, \bm{q}_{i,n}) \in SO(3)^n$. When combined with the Euclidean norms, the quaternion metric and the more natural geodesic metric are not functionally equivalent and hence yield, in general, different nearest neighbors. For small distances, i.e., for high sampling, the metrices are asymptotically identical.

To test whether this choice of metrics impacts the accuracy of the MI results, we compared our choice to the composite metric using $d_\text{quat}$ and the maximum-norm in $\mathbb{R}^2$, which is functionally equivalent to the geodesic composite metric but slightly less efficient to evaluate than $d_{\text{quat}^2}$. For $10^5$ frames, no significant difference between the MI values was seen.

\subsubsection{Volumes of balls in $SO(3)^n$}
\label{subsubsec:volumes_of_balls_in_SO3}

The volumes $V(r) = \int_{d(\bm{q}_i,\bm{y}) < r} d\bm{y}$ (dark green in Figure~\ref{fig:explanation_fig}B), enclosed by the kNN radius $r$, read
\begin{align*}
    V_1(r) &= 8\pi (r' - \sin r'), \\
            r'  &= 2 \arccos{\left( 1 - \frac{r^2}{2} \right)},
\end{align*}
for $d_\text{quat}$ in $SO(3)$. The respective volumes for $d_{\text{quat}^2}$ (in $SO(3)^2$) and $d_{\text{quat}^3}$ (in $SO(3)^3$) reduce to  
\begin{align*}
    V_2(r) &= 2^{10} \pi^2 \iint\displaylimits_{\mathcal{V}_2} \sin^2 \phi_A \sin^2 \phi_B d\phi_A d\phi_B, \\
    V_3(r) &= 2^{15} \pi^3 \\ 
           & \iiint\displaylimits_{\mathcal{V}_3} \sin^2 \phi_A \sin^2 \phi_B \sin^2 \phi_C d\phi_A d\phi_B d\phi_C,
\end{align*}
respectively, with $\mathcal{V}_2 = \{ 2-\cos \phi_A-\cos \phi_B \leq \frac{r^2}{2} \} \cap \{ 0 \leq \phi_A,\phi_B \leq \frac{\pi}{2} \}$ and $\mathcal{V}_3 = \{ 3-\cos \phi_A-\cos \phi_B-\cos \phi_C \leq \frac{r^2}{2} \} \cap \{ 0 \leq \phi_A,\phi_B,\phi_C \leq \frac{\pi}{2} \}$. The integrals were solved numerically for $10^4$ equally spaced values of $r$ using the software Mathematica 10.0\cite{Mathematica100} and the multidimensional rule; the results were stored in a lookup table. Cubic interpolation was used to obtain results from the stored values.

\section{Methods}

\subsection{Nearest-neighbor search}
\label{subsec:nearest_neighbor_search}

Nearest-neighbor searches were performed using the Non-Metric Space Library 1.7.3.6\cite{nmslib_2013} (NMSLIB)\footnote{\url{https://github.com/nmslib/nmslib}} and the above metrics. Each data set was indexed in a vantage-point tree\cite{Uhlmann_1991_balltree, Yianilos_1993_balltree} (VP-tree) that rests on the triangle inequality. Our version of the NMSLIB, modified to include the orientational metrices, is available online\footnote{\url{https://gitlab.gwdg.de/lheinz/nmslib_quaternion}}.

\subsection{Accuracy assessment}

\subsubsection{Test distributions}
\label{subsubsec:test_distributions}

To assess the accuracy of our method, we used analytical test-distributions $p^{(\mu)}$ in $SO(3)^1$, $SO(3)^2$, and $SO(3)^3$, derived from 
\begin{equation*}
    p^{(\mu)}_1(\bm{q}) = \frac{1}{Z^{(\mu)}} \cos^\mu \phi_1 = \frac{1}{Z^{(\mu)}} q_1^\mu,
\end{equation*}
with a quaternion $\bm{q} \in SO(3)^1$, the first quaternion component $q_1$, the first azimuthal angle in spherical coordinates for the 3-sphere $\phi_1 \in [0,\pi/2)$, and the appropriate normalization constant $Z^{(\mu)}$ (Figure~\ref{fig:Si_and_MI}A). The analytical expression for the configurational entropy $\int d\bm{q}  p^{(\mu)}_1 \log  p^{(\mu)}_1$ reads 
\begin{align*}
    S_1^{(\mu)} &= \frac{1}{2} \left\{ \mu \psi\left(\frac{\mu+4}{2} \right) -\mu \psi\left(\frac{\mu+1}{2}\right) \right. \\
    &\left. + 2 \log \left(\frac{\Gamma(\frac{\mu+1}{2})}{\Gamma(\frac{\mu+4}{2})} \right) + \log (64\pi^3) \right\},
\end{align*}
using the gamma function $\Gamma$ and $k_B = h = 1$ for simpler notation. As illustrated in Figure~\ref{fig:Si_and_MI}A, the distribution depends on the localization parameter $\mu$; a value of $0$ yields a uniform distribution; larger $\mu$ values yield increasingly narrower distributions.

For $(\bm{q}_1,\bm{q}_2) \in SO(3)^2$ and $(\bm{q}_1,\bm{q}_2,\bm{q}_3) \in SO(3)^3$, probability distributions $p^{(\mu)}_2((\bm{q}_1,\bm{q}_2)) = p^{(\mu)}(\bm{q}_1) p^{(\mu)}(\bm{q}_2)$ and $p^{(\mu)}_3((\bm{q}_1,\bm{q}_2,\bm{q}_3)) = p^{(\mu)}(\bm{q}_1) p^{(\mu)}(\bm{q}_2) p^{(\mu)}(\bm{q}_3)$ were used to obtain uncorrelated distributions with entropies $S_2^{(\mu)} = 2 S_1^{(\mu)}$, and $S_3^{(\mu)} = 3 S_1^{(\mu)}$, respectively.

To also assess the accuracy for correlated distributions with $(\bm{q}_1,\bm{q}_2) \in SO(3)^2$, the test-distribution
\begin{align*}
    p_{2,\text{corr}}^{(\mu)}((\bm{q}_1,\bm{q}_2)) &= \frac{1}{8\pi^2 Z^{(\mu)}} \cos^\mu \left( d_\text{geo}(\bm{q}_1,\bm{q}_2) \right) \\
    &=  \frac{1}{8\pi^2 Z^{(\mu)}} | \bm{q}_1\cdot \bm{q}_2 |^\mu
\end{align*}
was used, which was designed such that the marginals with respect to $\bm{q}_1$ and $\bm{q}_2$ are $p^{(\mu)}_1(\bm{q}_2)$ and $p^{(\mu)}_1(\bm{q}_1)$, respectively.
The localization $\mu$ here controls the degree of correlation between $\bm{q}_1$ and $\bm{q}_2$, ranging from an uncorrelated uniform distribution ($\mu = 0$) on $SO(3)^2$ to strongly correlated distributions for lager values. The entropy of this distribution is
\begin{equation*}
    S_{2,\text{corr}}^{(\mu)} = S_1^{(\mu)} + \log(8\pi^2),
\end{equation*}
where $\log(8\pi^2)$ is the entropy of a free rotor.

Samples were obtained using a rejection method: First, a random point in $\bm{Q} = (\bm{q}_1, \hdots, \bm{q}_n) \in SO(3)^n$ was drawn from a uniform distribution by drawing $n$ quaternions from the uniform distribution on the 3-sphere. Next, a random number $a$ was drawn from a uniform distribution between $0$ and $\max(p_n^{(\mu)})$. $\bm{Q}$ was accepted if $a < p_n^{(\mu)}(\bm{Q})$ and was rejected otherwise. This process was repeated until the desired number of samples was obtained.

The accuracy of our method was assessed for each test distribution for localization parameters between $\mu=0$ and $50$, nearest-neighbor $k$-values of $1,5,9$, and $13$, and with $10^2$ to $10^5$ frames ($n_f$). The computed entropy and MI values were compared to the analytical results. To obtain statistical error estimates, the calculations for each parameterset was repeated $1000$ times.

\subsection{Molecular dynamics simulations}
\label{subsec:MD_simulations}

All MD simulations were carried out using a modified version\footnote{\url{https://github.com/Tsjerk/gromacs/tree/rtc2018}} of the software package Gromacs 2018\cite{Gromacsa,Gromacsb,Gromacs4,Gromacs4.5,Gromacs2014} with an additional center of mass motion (COM) removal\cite{RTC} method, used to individually constrain all oxygen atoms. We furthermore made small additional changes to apply COM removal to individual atoms and to overcome the limit of $254$ COM removal groups\footnote{\url{https://gitlab.gwdg.de/lheinz/gromacs-rtc2018_modif}}.
The CHARMM36m force field\cite{charmm1,Karplus,charmm2,charmm36,charmm36m} and the CHARMM-TIP3P water model\cite{TIP3P} were used. All water molecules were subjected to SETTLE\cite{SETTLE} constraints (i.e., rigid), and the leap frog integrator with a time step of $2\,$fs was used. Electrostatic forces were calculated using the Particle-Mesh Ewald (PME) method\cite{PME} with a $1.2\,$nm real space cut-off; the same cut-off was used for Lennard-Jones potentials\cite{LJ_potential}. In all simulations, the V-rescale thermostat\cite{V_rescale} with a time constant of $0.1\,$ps, and, if applicable, the Parrinello-Rahman barostat\cite{Andersen_barostat, Parrinello_Rahman_barostat} with a time constant of $1.0\,$ps and $1\,$bar pressure were used.

A total of $1728$ water molecules were placed within a cubic simulation box, and the system was equilibrated for $1\,$ns at $300\,$K as a NPT ensemble. From the equilibrated system, resulting in a box size of approximately $3.7\,$nm, three $1\,\mu$s production runs were started, as shown in the first column of Figure~\ref{fig:MD_results}. Run \textbf{m} ("mobile") was carried out as described above. To benchmark our method against the established method of thermodynamic integration (TI), a system with only rotational degrees of freedom was constructed. To this end, all oxygens were position-constrained using COM removal as shown in the first column of Figure~\ref{fig:MD_results} in run \textbf{p} ("pinned"), allowing only rotational movements around the oxygen atom under NVT conditions. The temperature was increased to $600\,$K, since the water molecules formed an almost rigid, ice-like hydrogen bond network at $300\,$K, showing only very little dynamics. Run \textbf{sp} ("sliced \& pinned") was simulated like \textbf{p}, but all water molecules within a slice of $0.5\,$nm width were removed to create a water-vacuum interface.

\subsubsection{Entropy calculation}
\label{subsubsec:entropy_calculation}

For all three test systems, the entropy of rotation was calculated as described in section~\ref{sec:theory}, each using a $1\,\mu$s trajectory with $10^5$ frames. 
For the MI terms, a cut-off depending on the distance between average molecule positions was used. Whereas including the MIs of many molecule pairs by using a large cut-off distance gave rise to a more accurate MIE, it also introduced larger noise due to limited sampling. For pairwise MI terms, the cut-off was chosen as $1.0\,$nm, because for larger distances, the MI terms vanished within statistical errors (see Figure~\ref{fig:MD_results}B and D). Similarly, triple MI contributions were cut off at $0.45\,$nm. 

Because the water molecules in system \textbf{m} were mobile, average positions across the obtained trajectory were unusable to define a cut-off. Therefore, the water molecules were relabeled in each frame, such that they remained as close as possible to a simple-cubic reference structure using permutation reduction\cite{Reinhard_2007_permute, Reinhard_2009_g_permute}, which left the physics of the system unchanged. In systems \textbf{p} and \textbf{sp}, the molecules were immobilized and the oxygen positions where used for applying the cut-off.

To quantify the precision of the method, the MD simulations and the subsequent entropy analyses were repeated in 10 independent calculations.

\subsubsection{Thermodynamic integration reference}
\label{subsubsec:TI_reference}

Reference entropy values for systems \textbf{p} and \textbf{sp} were obtained using thermodynamic integration\cite{Kirkwood_1935_TI,Peter_2004_TI,Reinhard_2007_permute} (TI). Interactions between water molecules were gradually switched off in a stepwise fashion to obtain the entropy difference between real water and non-interacting water. The absolute rotational entropy was obtained as the sum of the excess entropy, obtained via TI, and the ideal gas contribution,
\begin{equation*}
    \frac{3Nk_B}{2} \log \left[ (4\pi^2)^\frac{2}{3} \cdot \frac{e k_B T}{2\pi\hbar^2} \prod_{i=1}^3 I_i^\frac{1}{3} \right],
\end{equation*}
where $I_i$ are the eigenvalues of the moment of inertia tensor of a water molecule. 

Both TI calculations were performed using the soft-core\cite{soft_core} parameters $\alpha=0.5$ and $\sigma=0.3$. Coulomb interactions were linearly switched off in $80$ windows of $20\,$ns each and further $10$ windows were used to subsequently switch off the van-der-Waals interactions. The first nanosecond of each window was discarded.

\section{Results and discussion}

\subsection{Test distributions}

\begin{figure*}
    \centering
    \includegraphics[scale=1]{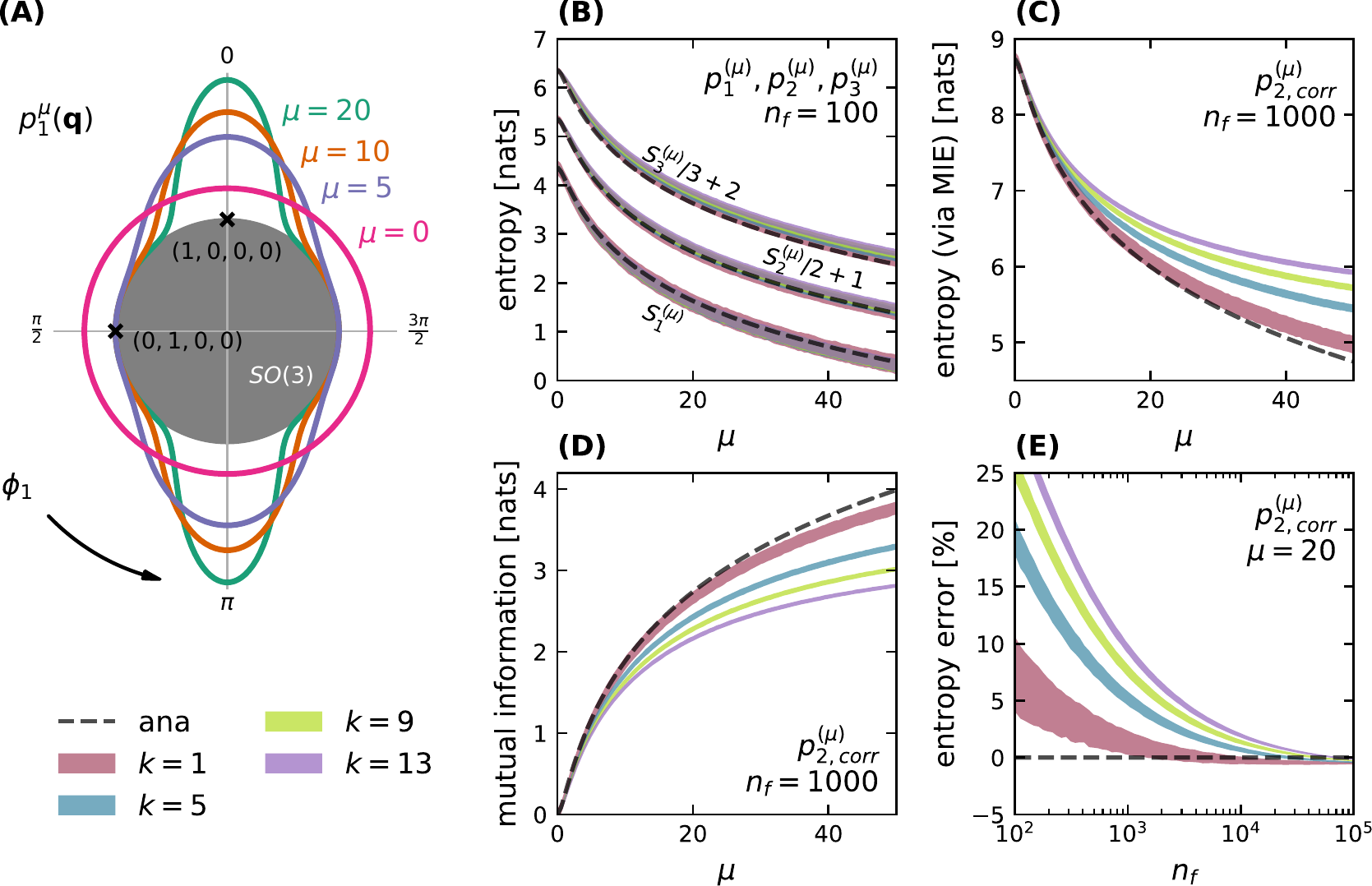}
    \caption{ Analytic test distribution compared to entropies and MI values obtained from density estimates. Panel \textbf{(A)} shows the distribution $p_{1}^{(\mu)}$ for increasing localizations $\mu$, illustrated by different colors. Here, we represent $SO(3)$ as a 1-sphere and the distribution is renormalized accordingly for this 1d representation. The north pole $(1,0,0,0)$ and the quaternion $(0,1,0,0)$ are indicated with black crosses. Panels \textbf{(B)}, \textbf{(C)}, and \textbf{(D)} show entropies and MI values obtained using the test distributions $p_{1}^{(\mu)},p_{2}^{(\mu)},p_{2,\text{corr}}^{(\mu)}$, and $p_{3}^{(\mu)}$ for varying coupling parameters $\mu$. Panel \textbf{(E)} shows the convergence of the results for increasing sample sizes. In panels \textbf{(B)} to \textbf{(E)}, the analytical result is shown by a dashed line; results for different $k$-values are colored according to the legend at the bottom left of the figure. Values that were fixed during the calculation, such as the choice of the test distribution, the number of frames $n_f$, or the coupling parameter $\mu$, are stated in a corner of the respective panel. The shown errors denote $1\sigma$ regions.}
    \label{fig:Si_and_MI}
\end{figure*}

We first assessed the accuracy of our method for three uncorrelated analytical test distributions (defined in $SO(3)^1$, $SO(3)^2$, and $SO(3)^3$) and one correlated analytical test distribution (defined in $SO(3)^2$), as described in section~\ref{subsubsec:test_distributions}. The distributions depend on the localization parameter $\mu$, which, for the uncorrelated distributions $p_{1}^{(\mu)},p_{2}^{(\mu)}$, and $p_{3}^{(\mu)}$, determines their width, as demonstrated in Figure~\ref{fig:Si_and_MI}A, and, for $p_{2,\text{corr}}^{(\mu)}$, controls the strength of the correlation. 

As can be seen in Figure~\ref{fig:Si_and_MI}B, the kNN estimator largely agrees with the analytic results (dashed lines) for the uncorrelated distributions $p_{1}^{(\mu)},p_{2}^{(\mu)}$, and $p_{3}^{(\mu)}$ for $\mu$ between $0$ (uniform distribution) and $50$ (strongly peaked), and the tested $k$-values between $1$ and $13$. The graphs for the three distributions are scaled and offset as indicated in the figure. We find that, for distributions $p_{1}^{(\mu)}$ and $p_{2}^{(\mu)}$, our method accurately reproduces the true entropy for all tested $\mu$-values within statistical errors, even for the small number of $100$ frames. The statistical errors amount to $0.25\,$nats (natural units of information) for $p_{2}^{(\mu)}$, $k=1$, $\mu=50$ or less. Also for $p_{3}^{(\mu)}$ and $k=1$, the analytical result is matched within statistical errors ($0.28\,$nats at $\mu=50$ at maximum), whereas larger values of $k$ lead to overestimated entropies of up to $0.7\,$nats ($k=13$, $\mu=50$), caused by the limited sampling of just $100$ frames and the increased dimensionality of $SO(3)^3$ compared to the other tested distributions.

Next, we assessed the accuracy for the correlated test distribution. The panels of Figure~\ref{fig:Si_and_MI}C and D show the entropy (calculated via the MIE as defined in eq \ref{eq:def_MI2}) and the MI of $p_{2,\text{corr}}^{(\mu)}$ for $1000$ frames, respectively. For the uniform distribution ($\mu=0$), the algorithm yields the analytic values of $2 \log(8\pi^2)$ and $0$ for entropy and mutual information, respectively. With increasing correlation $\mu$, the entropy is increasingly overestimated as MI is underestimated. Both effects are more pronounced for larger $k$-values: Whereas for $k=1$, the algorithm yields accurate values within statistical errors up to a correlation of $\mu \approx 20$, the results deviate significantly for $k=13$ even for very small $\mu$-values. Overall, small $k$-values, such as $k=1$, yield high accuracy but with reduced precision (i.e., larger statistical errors) compared to large $k$-values like $13$, which gives rise to smaller statistical errors but reduced accuracy. 

To further assess this trade-off and the convergence properties of our method, we calculated the relative entropy errors for $p_{2,\text{corr}}^{(20)}$ for sampling between $10^2$ and $10^5$ frames, shown in Figure~\ref{fig:Si_and_MI}E. For $k=1$ and only $100$ frames, the method overestimates the true entropy by $5$ to $10\%$, which quickly drops to below $1\%$ for more than $2\cdot10^3$ frames. For larger $k$-values, the entropy errors increase and the convergence becomes slower, e.g., $k=13$ requires $2\cdot10^4$ frames to achieve an entropy error of less than $1\,$\%. The statistical errors at $10^5$ frames are $0.11\,$\% and $0.05\,$\% for $k$-values of $1$ and $13$, respectively. Overall, $k=1$ yields somewhat lower precision but significantly faster convergence compared to larger values, which becomes even more pronounced in higher dimensions. We therefore consider this value the optimal choice for the systems at hand and used it for all subsequent analyses. 

The kNN entropy estimator rests on the assumption that the density is approximately constant and isotropic within each $k$-nearest-neighbor ball (see Figure~\ref{fig:explanation_fig}B). This assumption implies that features of the true distribution that are smaller than the average distance between sample points are not resolved, which, in case of poor sampling, inevitably leads to an overestimated entropy, as seen for $p_{3}^{(\mu)}$ with large $k$ or as shown in Figure~\ref{fig:Si_and_MI}E. The assumption of isotropy no longer holds for highly correlated datasets, such as $p_{2,\text{corr}}^{(\mu)}$ for large values of $\mu$. In this case, also the $k$-nearest neighbors to each sample point are correlated and thus not isotropically distributed, which is not reflected by an isotropic kernel, i.e., a ball. For Euclidean spaces, this problem was addressed by using anisotropic kernels\cite{Hensen_2009_anisotripic, Gao_2015_mutual_information}. Although this idea could also be applied in $SO(3)^n$, the correlation of water molecules at standard conditions is weak enough (Figure~\ref{fig:MD_results}A) to allow for sufficiently accurate results under the isotropy assumption. 

The trade-off between accuracy and precision with respect to the $k$-value is a general property of kNN entropy estimators, which has been characterized previously\cite{Khan_2007_mutual_information_estimation, Gao_2015_mutual_information}, and is intuitively accessible: Whereas averaging over an increasing number of neighbors reduces statistical uncertainties and thus improves precision, the assumptions of approximately constant isotropic densities are applied to increasingly larger balls, resulting in increasingly overestimated entropies for distributions with small scale features or strong correlations.

Overall, the kNN method with $k=1$ yields most accurate results while being only slightly less precise than estimators with lager $k$. It retrieves the analytical entropies within statistical errors for the uncorrelated distributions, as well as for the correlated distribution with $\mu<20$ using just $100$ and $1000$ frames, respectively. 

\subsection{Entropy calculated from MD simulations}

\begin{figure*}
    \centering
    \includegraphics[scale=1]{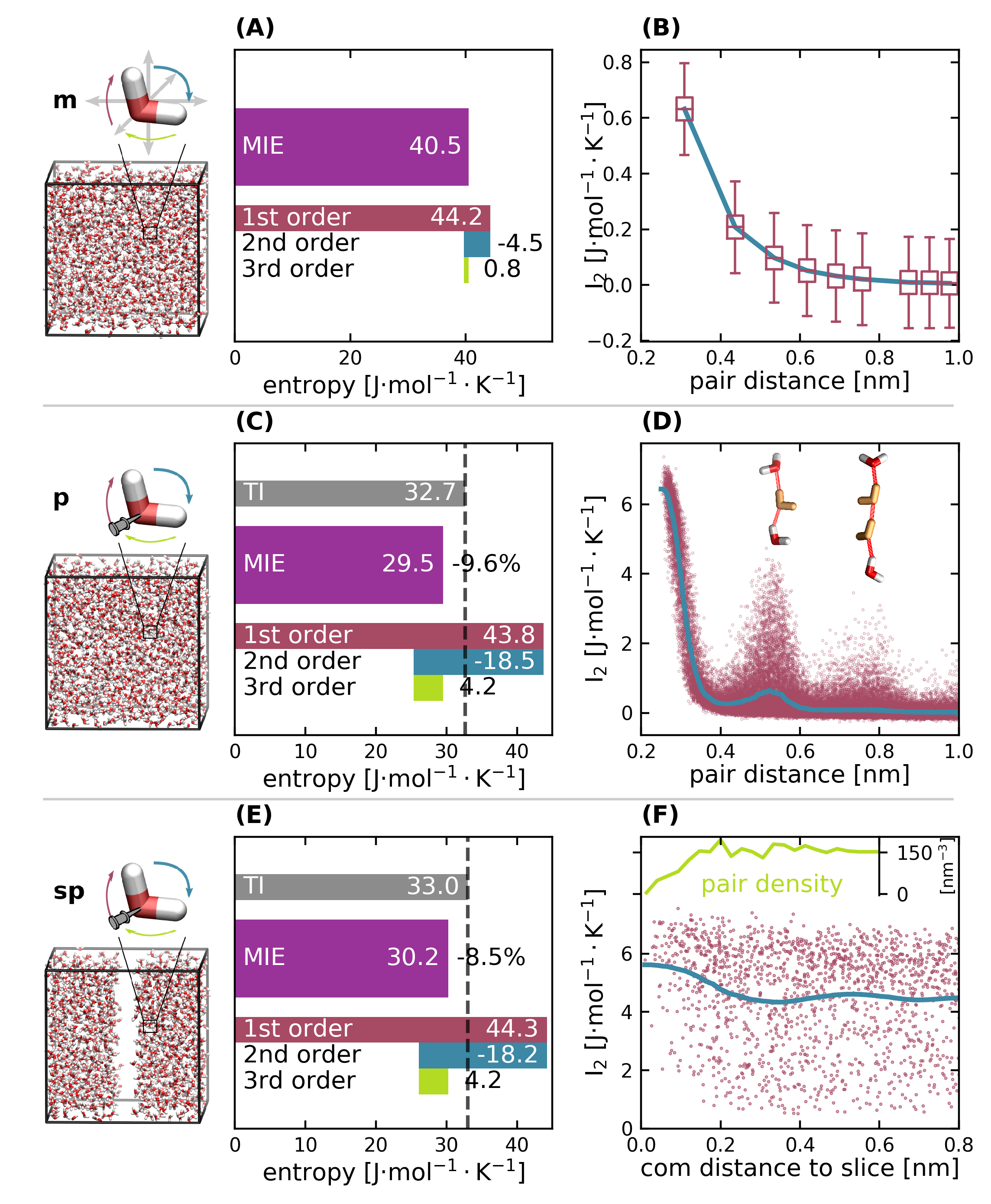}
    \caption{Entropies and MI contributions for the systems \textbf{m}, \textbf{p}, and \textbf{sp}. The first column shows the three considered MD systems. Panels (\textbf{A}), (\textbf{C}), and (\textbf{E}) show the rotational entropy computed using the MIE in purple; its breakdown into contributions by 1st to 3rd order is visualized underneath. For systems \textbf{p} and \textbf{sp}, the result is drawn in comparison to the TI values in gray. Panels (\textbf{B}) and (\textbf{D}) show the mutual information $I_2$ between all considered pairs of water molecules depending on their distance. The blue lines correspond to running Gaussian averages. Panel (\textbf{F}) displays $I_2$ between pairs of molecules that are closer than $0.33\,$nm in relation to the their distance to the vacuum slice in system \textbf{sp}. The inset in green shows the molecule pair density with respect to the center of mass distance to the slice.}
    \label{fig:MD_results}
\end{figure*}

Having assessed our rotational entropy method against analytic test distributions, we tested its accuracy for more realistic systems of up to $1728$ interacting water molecules. To this end, we simulated three atomistic water MD systems (Figure~\ref{fig:MD_results}, left column), as described in section~\ref{subsec:MD_simulations}. For all systems, 10 independent MD simulations were performed, and for each system, entropies were calculated via a MIE as explained in section~\ref{subsubsec:entropy_calculation}. 



System \textbf{m} ("mobile") comprises 1728 unconstrained water molecules. As shown in Figure~\ref{fig:MD_results}A, an absolute rotational entropy of $(40.53 \pm 0.04)\,$\JmolK{} per molecule is obtained, to which the first, second, and third MI orders contribute $(44.2349 \pm 0.0007)\,$\JmolK{}, $(-4.550 \pm 0.015)\,$\JmolK{}, and $(0.85 \pm 0.04)\,$\JmolK{}, respectively. Note that the provided values are averages and standard deviations of the 10 independent calculations and that the uncertainties are too small to be shown as error bars in Figure~\ref{fig:MD_results}. The pair-mutual information terms $I_2$, shown in Figure~\ref{fig:MD_results}B, reach a maximum of $0.8\,$\JmolK{} for very close water molecules and vanish monotonically for molecules that are, after permutation reduction and on average, separated by more than $\approx 0.8\,$nm. Note that the discrete nature of distances in Figure~\ref{fig:MD_results}B is due to the choice of a simple cubic reference structure for permutation reduction.

To compare the obtained absolute entropies to TI\cite{Kirkwood_1935_TI,Peter_2004_TI} (described in section~\ref{subsubsec:TI_reference}), the water movement was restricted to the rotational degrees of freedom in system \textbf{p} ("pinned") by pinning each molecule as described in section~\ref{subsec:MD_simulations}. Here, the rotational entropy, shown in panel C, is reduced to $(29.53 \pm 0.03)\,$\JmolK{}. The 2nd and 3rd order mutual information terms contribute $(-18.47 \pm 0.01)\,$\JmolK{} and $(4.21 \pm 0.02)\,$\JmolK{}, respectively. Compared to the results from TI shown in gray, the entropy is underestimated by $9.6\,$\% due to the limited sampling of the strongly correlated system. Similar to what we observe for the analytical test case depicted in Figure~\ref{fig:Si_and_MI}D, the MI terms are underestimated for strong correlations, of which the 3rd order is most severely affected due to the high dimensionality of the sampling space.

The $I_2$ terms, illustrated in Figure~\ref{fig:MD_results}D, show a maximum of $7\,$\JmolK{} and indicate that water molecules decorrelate beyond $\approx 0.4\,$nm. The distribution shows secondary and tertiary peaks around $0.55\,$nm and $0.80\,$nm that arise from indirect coupling via one or two mediating water molecules, as indicated by the structures shown in Figure~\ref{fig:MD_results}D. In this case, the correlations between the molecule pairs are not due to direct interactions; instead, mediating water molecules (orange) enhance distant orientation correlations via short hydrogen-bonded chains (shown in red). This finding demonstrates that the method is able to identify regions of locally coupled water molecules and to quantify the resulting entropy losses, thus providing a spatially resolved picture of entropy changes.

To further assess and demonstrate the accuracy of the method for systems with spatial features, we included a $0.5\,$nm vacuum slice in system \textbf{sp} ("sliced \& pinned", Figure~\ref{fig:MD_results}), such that the dynamics of water molecules at the surface differs from those molecules in the bulk. For system \textbf{sp}, the accuracy of our entropy estimation relative to TI improves to $8.5\,$\%, whereas the contributions by higher MI orders remain almost identical (see Figure~\ref{fig:MD_results}E). We assume that the improved accuracy is due to the smaller number of molecules (1728 vs. 1493 with slice) and possibly because the vacuum slice limits the range of many-particle correlations that would not be captured by a 3rd order approximation.

Figure~\ref{fig:MD_results}F shows the $I_2$ terms of molecule pairs that are closer than $0.33\,$nm, i.e., those that are within their first hydration shells, relative to their distance to the slice. The correlations of pairs that are close to the vacuum interface are increased to $5.6\,$\JmolK{} on average compared to $4.1\,$\JmolK{} in bulk. Although the entropy per molecule increases compared to system \textbf{p}, mainly due to the dominating 1st order term (see Figure~\ref{fig:MD_results}C~and~E), the increased correlations at the surface and their associated entropy losses contribute to the thermodynamic unfavorability of water at a (hydrophobic) vacuum interface. 



The MIE approaches the TI values for systems \textbf{p} and \textbf{sp} to $9.6\,$\% and $8.5\,$\%, respectively, and additionally yields information about individual correlations and their associated entropy losses, thus providing spatial resolution. Remarkably, about 25-fold less computer time was required for the MIE compared to TI for the shown examples. 

The large 2nd and 3rd order contributions, illustrated in Figure~\ref{fig:MD_results}C~and~E, show that both systems with pinned water exhibit strong correlations between water molecules. As for the test distributions illustrated in Figure~\ref{fig:Si_and_MI}, strong correlations result in systematically underestimated MI values. Due to their high dimensionality, and thus low sampling density, we expect the 3rd order MIE contributions for systems \textbf{p} and \textbf{sp} to be mostly affected, contributing to their overall underestimated entropy. For the same reason, we expect entropies calculated from more loosely coupled mobile water to yield markedly more accurate results. 

Although a direct comparison to TI is impossible for system \textbf{m}, we expect that the errors due to the truncation of higher order MI terms, observed for the more tightly correlated systems \textbf{p} and \textbf{sp}, are larger than for unconstrained water. Therefore, the approximation of the truncated MIE yields more accurate results for realistic solute systems. These two effects combined, the performances obtained for the more correlated pinned water systems provide upper bounds for the expected errors.

\section{Conclusion}

We developed an estimator for spatially resolved rotational solvent entropies based on a truncated mutual information expansion and the $k$-nearest-neighbor algorithm on $SO(3)^n$. Accuracy and computational efficiency were assessed for both analytical test distributions and for systems of up to 1728 water molecules, described by atomistic MD simulations.

For the uncorrelated test distributions in $SO(3)^1$, $SO(3)^2$, and $SO(3)^3$, the estimator with $k=1$ yields accurate entropies for as little as $100$ sample points. For the correlated test distribution $p_2^{(\mu)}$, the entropies are overestimated for increasing coupling, caused by underestimating mutual information terms. The latter effect is especially pronounced for large $k$-values. Precision increased only marginally for larger $k$ at the cost of decreased accuracy, which led us to conclude that $k=1$ represents the best trade-off for the problem at hand. 
We furthermore demonstrated convergence within $2\cdot10^3$ frames for a correlated distribution ($\mu=20$) and therefore expect our approach to accurately describe correlations of water molecules already in relatively short MD trajectories of $100\,$ns to $1\,\mu$s.

For the considered MD systems, we find agreement within $9.6\,$\% and $8.5\,$\% with TI for pinned waters in systems \textbf{p} and \textbf{sp}, respectively, corresponding to energy deviations ($-T\Delta S$) of $0.94\,$\kJmol{} and $0.84\,$\kJmol{} per water molecule at $300\,$K. The obtained rotational entropic contributions to the free energy are precise within $\pm 0.008\,$\kJmol{} and $\pm 0.018\,$\kJmol{}, respectively. For the binding of a small ligand that displaces $10$ water molecules at the binding pocket, we therefore expect to obtain absolute rotational entropy-contributions corresponding to an accuracy of at least $10\,$\kJmol{} and to resolve rotational entropy differences corresponding to at least $0.06\,$\kJmol{}. As seen in the second column of Figure~\ref{fig:MD_results}, fully mobile water exhibits considerably smaller correlations than pinned water, rendering the tests using pinned water a tough benchmark compared to realistic solute systems. For a protein/water system, we would therefore expect markedly smaller error margins.

The algorithm provides spatial resolution by assessing the mutual information contributions on the level of individual molecules, distinguishing it from, e.g., GIST\cite{Wallace_1987_IST,Baranyai_1989_IST, Lazaridis_1998A_IST, Lazaridis_1998B_IST,Nguyen_2012_GIST,Nguyen_2012_GIST_erratum}. For the hydrophobic vacuum interface, we calculated an entropy loss due to an increase in mutual information close to the surface. The ability to resolve the origin of entropy changes renders the method a promising tool to enhance our understanding of processes like the hydrophobic effect and the thermodynamics of solvated complex heterogeneous biomolecules in general.

Work on including the contributions by the translational entropy and the translation-rotation correlation to the overall entropy is in progress and will be published elsewhere. Also, our method can be extended to include intramolecular entropy contributions of flexible solvents, e.g., simulated water without SETTLE\cite{SETTLE} constraints. In this case, additional correlation terms would arise from pairwise correlations between the internal degrees of freedom, translation, and rotational, as well as the respective triple-correlation terms, which might be challenging to converge.

Although in this study we restricted the application and assessment of our approach to water, generalization to other solvents is straight forward. An implementation is available for download\footnote{\url{https://gitlab.gwdg.de/lheinz/hydration_entropy}} as a python module\cite{MDAnalysis1,MDAnalysis2} with a C++ backend for fast neighbor search.

\begin{acknowledgement}

L.P.H thanks the International Max Planck Research School for Physics of Biological and Complex Systems for support through a PhD Fellowship. Both authors thank Petra Kellers for proofreading the manuscript. 

\end{acknowledgement}


\providecommand{\latin}[1]{#1}
\makeatletter
\providecommand{\doi}
  {\begingroup\let\do\@makeother\dospecials
  \catcode`\{=1 \catcode`\}=2 \doi@aux}
\providecommand{\doi@aux}[1]{\endgroup\texttt{#1}}
\makeatother
\providecommand*\mcitethebibliography{\thebibliography}
\csname @ifundefined\endcsname{endmcitethebibliography}
  {\let\endmcitethebibliography\endthebibliography}{}

\end{document}